\documentclass[a4paper,12pt]{article}
\usepackage{epsfig}
\usepackage[dvips,usenames]{color}
\usepackage{graphicx}
\usepackage{color}
\usepackage{colortbl}

\newlength{\dinwidth}
\newlength{\dinmargin}
\newcommand{\spur}[1]{\not\! #1 \,}
\begin{document}
\title{Puzzles in $B\to \pi\pi,\pi K$ decays: Possible implications for
 R-parity violating supersymmetry}

\author{Ya-Dong Yang$^{1,}$\thanks{E-mail address:
yangyd@henannu.edu.cn }, Rumin Wang$^{1,2}$, and G.R.Lu$^{1,2}$
\\
{\footnotesize {$^1$ \it Department of Physics, Henan Normal
University, XinXiang, Henan 453007, P.R.China
 }}
 \\
 {\footnotesize {$^2$ \it Institute of Particle Physics, Huazhong Normal University,
  Wuhan, Hubei 430070, P.R.China }}
  }

\maketitle

\begin{abstract}
Recent experiments suggest that certain data of $B \to \pi\pi,\pi K$
decays are inconsistent with the standard model expectations. We try
to explain the discrepancies with R-parity violating suppersymmetry.
By employing the QCD factorization approach, we study these decays
in the minimal supersymmetric standard model with R-parity
violation. We show that R-parity violation can resolve the
discrepancies in both $B \to \pi\pi$ and $B \to \pi K$ decays, and
find that in some regions of parameter spaces  all these
requirements, including the $CP$ averaged branching ratios and the
direct $CP$ asymmetries, can be satisfied.  Furthermore, we have
derived stringent bounds on relevant R-parity violating couplings
from the latest experimental data, and some of these  constraints
are stronger than the existing bounds.

\end{abstract}

\vspace{1.5cm} \noindent {\bf PACS Numbers:  12.60.Jv,
  12.15.Mm, 12.38.Bx, 13.25.Hw}

\newpage
\section{Introduction}
The detailed study of B meson decays plays an essential role for
understanding the $CP$ violation and the physics of flavor. Recent
experimental measurements
\cite{belle1,be2,be3,be4,be5,be6,babar1,ba2,ba3,ba4,ba5,ba6} have
shown that some hadronic B decays to two pseudoscalar mesons deviate
from the standard model (SM) expectations .

In the $B \to \pi\pi$ decays, there are three such discrepancies:
the direct $CP$ asymmetry for the mode $B \to \pi^+\pi^-$ is very
large \cite{be2,ba4}, the $\pi^0\pi^0$ mode is found to have a much
larger branching ratio $(\approx 1.5 \times 10^{-6})$ \cite{be4,ba5}
than  the SM  expectations $(\sim 10^{-7})$
\cite{benekeNPB675,PQCD}, and the theoretical estimation of $B \to
\pi^+\pi^-$  branching ratio \cite{benekeNPB675,PQCD} are about 2
times larger than the current experimental average
\cite{belle1,babar1}. The ``$B \to \pi\pi$ puzzle" is reflected by
the following quantities \cite{burasNP}:
\begin{eqnarray*}
R^{\pi\pi}_{+-}&\equiv& 2\left[\frac{\mathcal{B}r(B^{+}\to
\pi^{+}\pi^0)+\mathcal{B}r(B^{-}\to
\pi^{-}\pi^0)}{\mathcal{B}r(B^0_d\to
\pi^{+}\pi^{-})+\mathcal{B}r(\bar{B}^0_d\to
\pi^{+}\pi^{-})}\right]\frac{\tau_{B_d}}{\tau_{B_u}}=(2.30\pm0.35)_{exp},\nonumber\\
R^{\pi\pi}_{00}&\equiv& 2\left[\frac{\mathcal{B}r(B^0_d\to
\pi^{0}\pi^0)+\mathcal{B}r(\bar{B}^0_d\to
\pi^{0}\pi^0)}{\mathcal{B}r(B^0_d\to
\pi^{+}\pi^{-})+\mathcal{B}r(\bar{B}^0_d\to
\pi^{+}\pi^{-})}\right]=(0.66\pm0.14)_{exp},
\end{eqnarray*}
here we use $\tau_{B_u}/\tau_{B_d}=1.069$ \cite{PDG2005}, the
central values calculated within the QCD factorization (QCDF) give
$R^{\pi\pi}_{+-}=1.24$ and $R^{\pi\pi}_{00}=0.07$
\cite{benekeNPB675}. In Ref. \cite{burasNP},
 Buras {\em et al.} pointed out these data would indicate the large nonfactorizable
contributions rather than  new physics (NP) effects, and  could be
perfectly accommodated in the SM. However,  it is hard to be
realized by explicit theoretical calculations.

The $B \to \pi K$ system consists of the four decay modes $B^0_d \to
\pi^{\pm} K^{\mp}, B^{\pm}_u \to \pi^{\pm} K, B^{\pm}_u \to \pi^{0}
K^{\pm}$ and $B^0_d \to \pi^0 K^0$, which are governed by QCD
penguin process $b \to sq\bar{q}$. The \textit{BABAR} and Belle
collaborations have measured the following ratios of the $CP$
averaged branching ratios \cite{burasratiopik}:
\begin{eqnarray*}
R&\equiv&\left[\frac{\mathcal{B}r(B^0_d\to
\pi^{-}K^{+})+\mathcal{B}r(\bar{B}^0_d\to
\pi^{+}K^{-})}{\mathcal{B}r(B^{+}\to
\pi^{+}K^0)+\mathcal{B}r(B^{-}\to
\pi^{-}\bar{K}^0)}\right]\frac{\tau_{B_u}}{\tau_{B_d}}=(0.79\pm0.06)_{exp},\\
R_c&\equiv&2 \left[\frac{\mathcal{B}r(B^+\to \pi^0
K^{+})+\mathcal{B}r(B^-\to \pi^0 K^{-})}{\mathcal{B}r(B^{+}\to
\pi^{+}K^0)+\mathcal{B}r(B^{-}\to
\pi^{-}\bar{K}^0)}\right]=(0.98\pm0.08)_{exp},\\
R_n&\equiv& \frac{1}{2}\left[\frac{\mathcal{B}r(B^0_d\to
\pi^{-}K^{+})+\mathcal{B}r(\bar{B}^0_d\to
\pi^{+}K^{-})}{\mathcal{B}r(B^0_d\to \pi^0
K^0)+\mathcal{B}r(\bar{B}^0_d\to \pi^0
\bar{K}^0)}\right]=(0.79\pm0.08)_{exp},
\end{eqnarray*}
where we have included the latest Belle \cite{be6} and
\textit{BABAR} \cite{ba6} measurements.  The ratio $R$, which is
expected to be only marginally affected by color-suppressed
electroweak (EW) penguins, does not show any anomalous behavior. The
``$B \to \pi K$ puzzle" is reflected by the  small value of $R_n$
which  is significantly lower than $R_c$.  Since $R_c$ and $R_n$
could be affected significantly by color-allowed EW penguins, the
``$B \to \pi K$ puzzle" may be a manifestation of NP in the EW
penguin sectors \cite{burasNP,gronauNP,mishima}, and will offer an
attractive avenue for physics beyond the SM to enter the $B \to \pi
K $ system \cite{Attractive avenue pik}.

Although these measurements represent quite a challenge for theory,
the SM is in no way ruled out yet since there are many theoretical
uncertainties in low energy QCD. The recent theoretical results for
$B \to \pi K$ \cite{lixn,lixqpp}  show that the next to leading
order corrections may be important. However, it will be under
considerable strain if the experimental data persist for a long
time.  Existence of NP as possible solutions have been discussed in
Refs. \cite{burasNP,gronauNP,datta}. Among those NP models that
survived EW data, one of the most respectable options is R-parity
violating (RPV) supersymmetry (SUSY). The possible appearance of RPV
couplings \cite{SUSY}, which will violate the lepton and baryon
number conservation, has gained full attention in searching for SUSY
\cite{report,allanach}. The effect of  RPV SUSY on B decays have
been extensively investigated previously in the literatures \cite{
RPVstudy, cskim, desh, yd, BtoVV,hexg}. In this work, we extend our
previous study of the $B \to VV$ decays \cite{BtoVV} to the $B \to
\pi\pi,\pi K$ decays using RPV SUSY theories by employing the QCD
factorization (QCDF) approach\cite{BBNS} for hadronic dynamics. At
the same theoretical ground, it would be interesting to know if we
could find solutions to the $B\to\pi \pi$ and  $B\to\pi K$ puzzles
besides the $B \to VV$ polarization puzzle \cite{BtoVV}. We show
that the $B \to \pi \pi,\pi K $ puzzles could be resolved in the
presence of the RPV couplings. Moreover, using the latest
experimental data and theoretical parameters, we try to explain all
available data including the $CP$ averaged branching ratios and the
direct $CP$ asymmetries by the relevant RPV couplings. We note that
the branching ratios of the $B \to \pi\pi, \pi K$ decays have also
been studied in RPV SUSY in \cite{hexg,RPVpipi}. In this study, we
present a new calculation of the decays with up-to-date inputs, such
as form factors, experimental measurements and so on.

The paper is arranged as follows.  In Sec. II, we calculate the
branching ratios and the $CP$ asymmetries of   $B\to \pi\pi, \pi K$
decays, which contain the SM contributions and the RPV effects using
the QCDF approach. In Sec. III, we tabulate theoretical inputs in
our analysis. Section IV is to deal with data and discussions, we
also display the allowed regions of the parameter space which
satisfy all the experimental data. Section V contains our summary
and conclusion.

\section{The theoretical frame for $B\to \pi\pi$ and $\pi K$ decays }

\subsection{ The decay amplitudes  in the SM }
  In the SM, the low energy effective Hamiltonian for
  the $\Delta B=1$ transition at the scale $\mu$ is given by \cite{coeff}
 \begin{eqnarray}
 \mathcal{H}^{SM}_{eff}&=&\frac{G_F}{\sqrt{2}}\sum_{p=u, c}
 \lambda_p \Biggl\{C_1Q_1^p+C_2Q_2^p
 +\sum_{i=3}^{10}\Big[C_iQ_i+C_{7\gamma}Q_{7\gamma}
 +C_{8g}Q_{8g}\Big] \Biggl\}+ h.c.,
 \end{eqnarray}
here  $\lambda_p=V_{pb}V_{pq}^* ~(p\in \{u,c\})$  and the detailed
definition of the operator base can be found in \cite{coeff}.

The decay amplitude for $B\to M_1M_2$ is
\begin{eqnarray}
  \mathcal{A}^{SM}(B\to M_1M_2)&=&\left< M_1M_2|{\cal H}^{SM}_{eff}|B \right> \nonumber\\
  &=&\frac{G_F}{\sqrt{2}}\sum_p \sum_i \lambda_p
  C_i(\mu)\left<M_1M_2|Q_i(\mu)|B\right>.
  \end{eqnarray}
The essential theoretical difficulty for obtaining the decay
amplitude arise from the  evaluation of hadronic matrix elements
$\langle M_1M_2|Q_i(\mu)|B\rangle$. There are at least three
approaches with different considerations to tackle the said
difficulty: the naive factorization (NF) \cite{NF1,NF2}, the
perturbative QCD \cite{PQCD}, and the QCDF \cite{BBNS}.

  The QCDF  \cite{BBNS} developed by Beneke, Buchalla, Neubert
  and Sachrajda is a
  powerful framework for studying charmless B decays.
  In Refs. \cite{benekeNPB675,ydpipipik,dsdupp},
  $B \to \pi\pi, \pi K$ decays have been analyzed in detail in the SM with the
QCDF approach.  We will also employe the  QCDF approach in this
paper.

The QCDF  \cite{BBNS} allows us to compute the nonfactorizable
corrections to the hadronic matrix elements $\langle M_1
M_2|O_i|B\rangle$ in the heavy quark limit. The  decay amplitude has
the form
\begin{eqnarray}
  \mathcal{A}^{SM}(B\to M_1M_2)
  =\frac{G_F}{\sqrt{2}}\sum_p \sum_i \lambda_p
  \Biggl\{a^p_i\langle M_2|J_2|0\rangle\langle
  M_1|J_1|B\rangle+b^p_i\langle M_1M_2|J_2|0\rangle\langle
  0|J_1|B\rangle\Biggl\},
  \end{eqnarray}
 where  the effective
parameters $a_i^p$  including nonfactorizable corrections at order
of $\alpha_s $.  They are calculated from the vertex corrections,
the hard spectator scattering, and the QCD penguin contributions.
The parameters $b_i^p$ are calculated from the weak annihilation
contributions.

\begin{figure}[h]
\begin{center}
\includegraphics[scale=0.9]{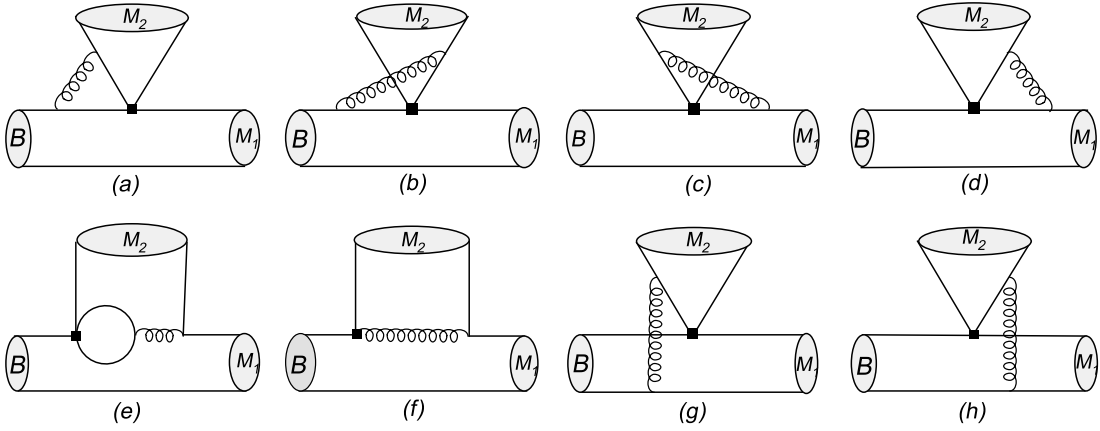}
\end{center}
\vspace{-0.6cm}
 \caption{\small The next to leading order nonfactorizable
  contributions to the coefficients $a^p_i$.}
 \label{NTL}
\end{figure}

Following Beneke and Neubert  \cite{benekeNPB675}, coefficients
 $a_i^p$ can be split into two parts:
$a_i^p=a_{i,I}^p+a_{i,II}^p$. The first part contains the NF
contribution and the sum of nonfactorizable  vertex and penguin
corrections, while the second one arises from the hard spectator
scattering. The coefficients read \cite{benekeNPB675}
 \begin{eqnarray}
&& a_{1,I} =C_1+ \frac{C_2}{N_C} \left[1+\frac{C_F \alpha_s}{4\pi}
V_{M_2}\right],\qquad \qquad \qquad \qquad a_{1,II} =
\frac{C_2}{N_C} \frac{C_F \alpha_s}{4\pi}
H_{M_2M_1},\nonumber\\
&& a_{2,I} =C_2+\frac{C_1}{N_C}\left[1+\frac{C_F \alpha_s}{4\pi}
V_{M_2}\right], \qquad \qquad \qquad \qquad a_{2,II} =
\frac{C_1}{N_C} \frac{C_F \alpha_s}{4\pi}
H_{M_2M_1},\nonumber\\
&&a_{3,I}=C_3+\frac{C_4}{N_C}\left[1+\frac{C_F \alpha_s}{4\pi}
V_{M_2}\right], \qquad \qquad \qquad \qquad a_{3,II} =
\frac{C_4}{N_C} \frac{C_F \alpha_s}{4\pi}
H_{M_2M_1},\nonumber\\
&&a_{4,I}^p =C_4+\frac{C_3}{N_C}\left[1+\frac{C_F \alpha_s}{4\pi}
V_{M_2}\right]+\frac{C_F \alpha_s}{4 \pi}\frac{P^p_{M_2,2}}{N_C},
\qquad ~ a_{4,II} = \frac{C_3}{N_C} \frac{C_F \alpha_s}{4\pi}
H_{M_2M_1},\nonumber\\
&&a_{5,I} = C_5+\frac{C_6}{N_C}\left[1+\frac{C_F \alpha_s}{4\pi}
(-12-V_{M_2}) \right], \qquad \qquad a_{5,II} = \frac{C_6}{N_C}
\frac{C_F \alpha_s}{4\pi}
(-H_{M_2M_1}),\nonumber\\
&&a_{6,I}^p=C_6+\frac{C_5}{N_C}\left[1-6\cdot\frac{C_F
\alpha_s}{4\pi}\right]+\frac{C_F \alpha_s}{4 \pi}
\frac{P^p_{M_2,3}}{N_C},\qquad ~ a_{6,II} = 0,\nonumber\\
&& a_{7,I}=C_7+\frac{C_8}{N_C}\left[1+\frac{C_F \alpha_s}{4\pi}
(-12-V_{M_2})\right], \qquad \qquad ~ a_{7,II} = \frac{C_8}{N_C}
\frac{C_F \alpha_s}{4\pi}
(-H_{M_2M_1}),\nonumber\\
&&a_{8,I}^p =C_8+\frac{C_7}{N_C}\left[1-6\cdot\frac{C_F
\alpha_s}{4\pi}\right]+\frac{\alpha_e}{9
\pi}\frac{P^{p,EW}_{M_2,3}}{N_C}, \qquad ~~~ a_{8,II} = 0,\nonumber\\
&&a_{9,I} =C_9+\frac{C_{10}}{N_C}\left[1+\frac{C_F \alpha_s}{4\pi}
V_{M_2}\right], \qquad \qquad \qquad \qquad a_{9,II} =
\frac{C_{10}}{N_C} \frac{C_F \alpha_s}{4\pi}
H_{M_2M_1},\nonumber\\
&&a_{10,I}^p=C_{10}+\frac{C_9}{N_C}\left[1+\frac{C_F \alpha_s}{4\pi}
V_{M_2}\right]+\frac{\alpha_e}{9 \pi}\frac{P^{p,EW}_{M_2,2}}{N_C},
\qquad a_{10,II} = \frac{C_9}{N_C} \frac{C_F \alpha_s}{4\pi}
H_{M_2M_1}, \label{coeffa}
 \end{eqnarray}
where $\alpha_s\equiv \alpha_s(\mu)$, $C_F=(N_C^2-1)/(2 N_C)$, and
$N_C=3$ is the number of colors. The quantities $V_{M}, H_{M_2M_1},
P^p_{M,2}, P^p_{M,3}, P^{p,EW}_{M,2}$ and $P^{p,EW}_{M,3}$ consist
of convolutions of hard-scattering kernels with meson distribution
amplitudes. Specifically, the terms $V_M$ come from the vertex
corrections in Fig. \ref{NTL}(a)-\ref{NTL}(d), $P^p_{M,2}$ and
$P^p_{M,3}~ (P^{p,EW}_{M,2}$ and $P^{p,EW}_{M,3})$ arise from QCD
(EW) penguin contractions and the contributions from the dipole
operators as depicted by Fig. \ref{NTL}(e) and \ref{NTL}(f).
$H_{M_2M_1}$ is due to the hard spectator scattering as Fig.
\ref{NTL}(g) and \ref{NTL}(h). For the penguin terms, the subscript
2 and 3 indicate the twist 2 and 3 distribution amplitudes of light
mesons, respectively.

In Eq.(\ref{coeffa}), $V_{M}$ $(M=\pi, K)$ contains  the
nonfactorizable  vertex corrections , which is
\begin{eqnarray}
 &&V_M=12\ln \frac{m_b}{\mu}-18+3 \int_0^1{\rm d}u
 \left(\frac{1-2u}{1-u}\ln u-i \pi\right)\Phi_{M}(u).
\end{eqnarray}

Next, the penguin contributions at  the twist-2 are described by the
functions
\begin{eqnarray}
P^p_{M,2}&=&C_1\left[\frac{2}{3}+\frac{4}{3}\ln
\frac{m_b}{\mu}-G_M(s_p)\right]+C_3\left[\frac{4}{3}+\frac{8}{3}\ln
\frac{m_b}{\mu}-G_M(0)-G_M(1)\right]\nonumber\\
&&+~(C_4+C_6)\left[\frac{4 n_f}{3}\ln
\frac{m_b}{\mu}-(n_f-2)G_M(0)-G_M(s_c)-G_M(1)\right]\nonumber\\
&&-~C^{eff}_{8g}\int_0^1 {\rm d}u ~\frac{2\Phi_{M} (u)}{1-u},\nonumber\\
P^{p,EW}_{M,2}&=& (C_1+N_CC_2)\left[\frac{2}{3}+\frac{4}{3}\ln
\frac{m_b}{\mu}-G_M(s_p)\right]-~C^{eff}_{7\gamma}\int_0^1 {\rm d}u
~\frac{3\Phi_{M} (u)}{1-u},
\end{eqnarray}
where $n_f = 5$ is the number of quark flavors, and $s_u = 0, s_c =
(m_c/m_b)^2$ are mass ratios involved in the evaluation of the
penguin diagrams. The function $G_M(s)$ is defined as
 \begin{eqnarray}
G_M(s)&=&-4\int_0^1{\rm d}u \int_0^1{\rm d}x~
x\bar{x}\ln{(s-x\bar{x}\bar{u} -i \epsilon)}\Phi_{M} (u).
 \end{eqnarray}

The twist-3 terms from the penguin diagrams are given by
\begin{eqnarray}
P^p_{M,3}&=&C_1\left[\frac{2}{3}+\frac{4}{3}\ln
\frac{m_b}{\mu}-\widehat{G}_M(s_p)\right]+C_3\left[\frac{4}{3}+\frac{8}{3}\ln
\frac{m_b}{\mu}-\widehat{G}_M(0)-\widehat{G}_M(1)\right]\nonumber\\
&&+~(C_4+C_6)\left[\frac{4 n_f}{3}\ln
\frac{m_b}{\mu}-(n_f-2)\widehat{G}_M(0)-\widehat{G}_M(s_c)
-\widehat{G}_M(1)\right]-~2 C^{eff}_{8g},\nonumber\\
P^{p,EW}_{M,3}&=& (C_1+N_CC_2)\left[\frac{2}{3}+\frac{4}{3}\ln
\frac{m_b}{\mu}-\widehat{G}_M(s_p)\right]-~3 C^{eff}_{7\gamma},
\end{eqnarray}
with
\begin{eqnarray}
\widehat{G}_M(s)&=&-4\int_0^1{\rm d}u \int_0^1{\rm d}x~
x\bar{x}\ln{(s-x\bar{x}\bar{u} -i \epsilon)}\Phi^{M}_p (u).
\end{eqnarray}

Finally, the hard spectator interactions can be written  as
\begin{eqnarray}
H_{M_2M_1}=\frac{4\pi^2}{N_C}\frac{f_B f_{M_1}}{m_B^2 f_0^{B \to
M_1}(0)}\int^1_0\frac{{\rm d}\xi}{\xi}\Phi_B(\xi)\int^1_0\frac{{\rm
d}u }{\bar{u}}\Phi_{M_2}(u)\int^1_0\frac{{\rm d}v
}{\bar{v}}\left[\Phi_{M_1}(v)+\frac{2\mu_{M_1}}{M_B}\Phi^{M_1}_p(v)\right].
\label{H}
\end{eqnarray}
Considering the off-shellness of the gluon in  hard scattering
kernel, it is natural to associate a scale $\mu_h \sim
\sqrt{\Lambda_{QCD} m_b}$ , rather than $\mu\sim m_b$.  For the
logarithmically divergent integral,
 we will parameterize it as in \cite{benekeNPB675}:
  $X_H=\int^1_0 {\rm d}u/u=-ln (\Lambda_{QCD}/m_b)+\varrho_He^{i\phi_H}$
$m_b/\Lambda_{QCD}$  with $(\varrho_H,\phi_H)$ related to the
contributions from hard spectator scattering. In the later numerical
analysis, we shall take $\Lambda_{QCD}=0.5GeV$,
$(\varrho_h,\phi_H)=(0,0)$ as our default values.

\begin{figure}[h]
\begin{center}
\includegraphics[scale=0.8]{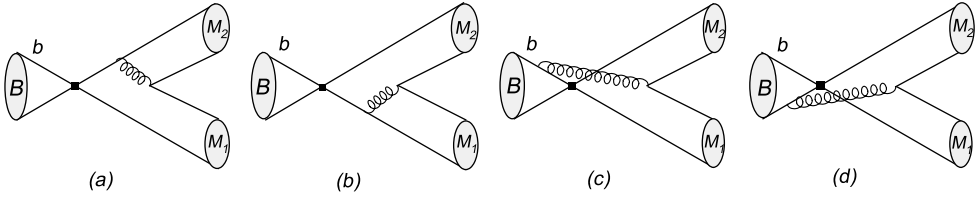}
\end{center}
\vspace{-0.6cm}
 \caption{\small The weak annihilation  contributions to the coefficients $b^p_i$.}
 \label{ANN}
\end{figure}

At leading order in $\alpha_s$, the annihilation contribution can
be calculated from the diagrams in Fig.\ref{ANN}. The annihilation
coefficients $(b_1, b_2), (b_3, b_4)$ and $(b^{EW}_3 , b^{EW}_4)$
correspond to the contributions of the tree, QCD penguins and EW
penguins operators insertions, respectively.  Using the asymptotic
light cone distribution amplitudes of the mesons, and assuming
SU(3) flavor symmetry, they can be expressed as
\begin{eqnarray}
&&b_1=\frac{C_F}{N_C^2}C_1A^i_1, \hspace{2cm}
b_3=\frac{C_F}{N_C^2}\left[C_3A^i_1+C_5\left(A^i_3+A^f_3\right)
+N_CC_6A^f_3\right],\nonumber\\
&&b_2=\frac{C_F}{N_C^2}C_2A^i_1, \hspace{2cm}
b_4=\frac{C_F}{N_C^2}\left[C_4A^i_1+C_6A^i_2\right],\nonumber\\
&&b_3^{EW}=\frac{C_F}{N_C^2}\left[C_9A^i_1+C_7\left(A^i_3+A^f_3\right)
+N_CC_8A^f_3\right],\nonumber\\
&&b_4^{EW}=\frac{C_F}{N_C^2}\left[C_{10}A^i_1+C_8A^i_2\right],
\label{coeffb}
\end{eqnarray}
and
\begin{eqnarray}
&&A^i_1\approx \pi \alpha_s \left[18 \left(
X_A-4+\frac{\pi^2}{3}\right)+ 2r^2_\chi X^2_A \right],\qquad ~
A^i_2=A^i_1, \qquad A^i_3=0,\nonumber\\
&&A^f_3\approx 12 \pi \alpha_s r_\chi \left(
2X^2_A-X_A\right),\qquad \qquad \qquad \qquad A^f_1=0, \qquad ~~
A^f_2=0. \label{An}
\end{eqnarray}
Here the superscripts i and f refer to gluon emission from the
initial and final state quarks, respectively. The subscript $k$ of
$A^{i,f}_k$ refers to one of the three possible Dirac structures
$\Gamma_1 \otimes \Gamma_2$, namely $k = 1$ for
$(V-A)\otimes(V-A)$, $k = 2$ for $(V-A)\otimes(V+A)$, and $k = 3$
for $(-2)(S-P)\otimes(S+P)$.  $X_A = \int^1_0 {\rm d}u/u$ is a
logarithmically divergent integral, and will be phenomenologically
parameterized in the calculation as $X_H$. As for the hard
spectator terms, we will evaluate the various quantities in
Eq.(\ref{An}) at the scale $\mu_h = \sqrt{\Lambda_{QCD} m_b}$.

With the coefficients in Eq.(\ref{coeffa}) and (\ref{coeffb}), we
can obtain the decay amplitudes of the SM part
$\mathcal{A}^{SM}_f$ (the subscript ``$f$" denotes the part of
QCDF contribution)
 and  $\mathcal{A}^{SM}_{a}$ (the subscript ``$a$" denotes the annihilation part).
 $B\to \pi \pi$ and $\pi K$ decay amplitudes are
given in Appendix A.

\subsection {R-parity violating SUSY effects in the decays}

In the most general superpotential of the minimal supersymmetric
Standard Model (MSSM), the RPV superpotential is given by
\cite{RPVSW}
\begin{eqnarray}
\mathcal{W}_{\spur{R}}&=&\mu_i\hat{L}_i\hat{H}_u+\frac{1}{2}\lambda_{[ij]k}
\hat{L}_i\hat{L}_j\hat{E}^c_k+
\lambda'_{ijk}\hat{L}_i\hat{Q}_j\hat{D}^c_k+\frac{1}{2}
\lambda''_{i[jk]}\hat{U}^c_i\hat{D}^c_j\hat{D}^c_k, \label{rpv}
\end{eqnarray}
where $\hat{L}$ and $\hat{Q}$ are the SU(2)-doublet lepton and
quark superfields and $\hat{E}^c$, $\hat{U}^c$ and $\hat{D}^c$ are
the singlet superfields, while i, j and k are generation indices
and $c$ denotes a charge conjugate field.

The bilinear RPV superpotential terms $\mu_i\hat{L}_i\hat{H}_u$
can be rotated away by suitable redefining the lepton and Higgs
superfields \cite{barbier}. However, the rotation will generate a
soft SUSY breaking bilinear term which would affect our
calculation through penguin level. However, the processes
discussed in this paper could be induced by tree-level RPV
couplings, so that we would neglect sub-leading RPV penguin
contributions in this study.

 The $\lambda$ and $\lambda'$
couplings in Eq.(\ref{rpv}) break the lepton number, while the
$\lambda''$ couplings break the baryon number. There are 27
$\lambda'_{ijk}$ couplings, 9 $\lambda_{ijk}$ and 9
$\lambda''_{ijk}$ couplings.  $\lambda_{[ij]k}$ are antisymmetric
with respect to their first two indices, and $\lambda''_{i[jk]}$
are antisymmetric with j and k. The antisymmetry of the  baryon
number violating couplings $\lambda''_{i[jk]}$ in the last two
indices implies that there are no $\lambda''_{ijk}$ operator
generating the $\bar{b}\rightarrow \bar{s}s \bar{s}$ and
$\bar{b}\rightarrow \bar{d} d \bar{d}$ transitions.

\begin{figure}[htbp]
\begin{center}
\begin{tabular}{c}
\includegraphics[scale=0.65]{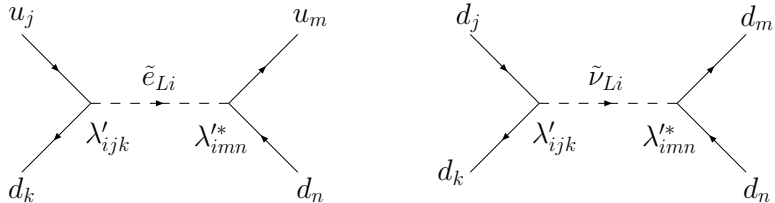}
\end{tabular}
\end{center}
\vspace{-0.6cm}
 \caption{\small Sleptons exchanging diagrams for nonleptonic B decays.}
 \label{Hp}
\end{figure}

\begin{figure}[htbp]
\begin{center}
\begin{tabular}{c}
\includegraphics[scale=0.65]{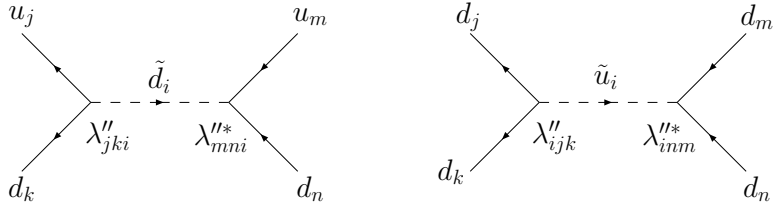}
\end{tabular}
\end{center}
\vspace{-0.6cm}
 \caption{\small Squarks exchanging diagrams for nonleptonic B decays.}
  \label{Hpp}
\end{figure}
From Eq.(\ref{rpv}), we can obtain the following four fermion
effective  Hamiltonian due to the sleptons exchange  as shown in
Fig.\ref{Hp}
\begin{eqnarray}
\mathcal{H}'^{\spur{R}}_{2u-2d}&=&\sum_i\frac{\lambda'_{ijm}
\lambda'^*_{ikl}}{2m^2_{\tilde{e}_{Li}}}
\eta^{-8/\beta_0}(\bar{d}_m\gamma^\mu P_Rd_l)_8(\bar{u}_k
\gamma_\mu P_Lu_j)_8,\nonumber\\
\mathcal{H}'^{\spur{R}}_{4d}&=&\sum_i
\frac{\lambda'_{ijm}\lambda'^*_{ikl}}{2m^2_{\tilde{\nu}_{Li}}}
\eta^{-8/\beta_0}(\bar{d}_m\gamma^\mu P_Rd_l)_8(\bar{d}_k\gamma_\mu
P_Ld_j)_8.
\end{eqnarray}
The four fermion effective Hamiltonian due to the squarks
exchanging as shown in Fig. \ref{Hpp} are
\begin{eqnarray}
\mathcal{H}''^{\spur{R}}_{2u-2d}&=&\sum_n\frac{\lambda''_{ikn}
\lambda''^*_{jln}}{2m^2_{\tilde{d}_{n}}}\eta^{-4/\beta_0}
\left\{\left[(\bar{u}_i\gamma^\mu P_Ru_j)_1(\bar{d}_k\gamma_\mu
P_Rd_l)_1-(\bar{u}_i\gamma^\mu P_Ru_j)_8(\bar{d}_k\gamma_\mu
P_Rd_l)_8\right]\right. \nonumber
\\&&\hspace{3cm}-\left[\left.(\bar{d}_k\gamma^\mu P_Ru_j)_1
(\bar{u}_i\gamma_\mu P_Rd_l)_1
-(\bar{d}_k\gamma^\mu P_Ru_j)_8
(\bar{u}_i\gamma_\mu P_Rd_l)_8\right]\right\},\nonumber\\
\mathcal{H}''^{\spur{R}}_{4d}&=&\sum_n
\frac{\lambda''_{nik}\lambda''^*_{njl}}{4m^2_{\tilde{u}_{n}}}
\eta^{-4/\beta_0}\left[(\bar{d}_i\gamma^\mu
P_Rd_j)_1(\bar{d}_k\gamma_\mu P_Rd_l)_1-(\bar{d}_i\gamma^\mu
P_Rd_j)_8(\bar{d}_k\gamma_\mu P_Rd_l)_8\right].
\end{eqnarray}
where $P_L=\frac{1-\gamma_5}{2},P_R=\frac{1+\gamma_5}{2},
\eta=\frac{\alpha_s(m_{\hat{f}_i})}{\alpha_s(m_b)}$ and
$\beta_0=11-\frac{2}{3}n_f$. The subscript for the currents
$(j_{\mu})_{1, 8} $ represents  the current in the color singlet
and octet, respectively.  The coefficients $\eta^{-4/\beta_0}$ and
$\eta^{-8/\beta_0}$ are due to the running from the sfermion mass
scale $m_{\hat{f}_i}$ (100 GeV assumed) down to the $m_b$ scale.
Since  it is always assumed in phenomenology for numerical display
that only one sfermion contributes at one time, we neglect the
mixing between the operators when we use the renormalization group
equation (RGE) to run $\mathcal{H}^{\spur{R}}$ down to the low
scale.

Generally, the  RPV couplings can  be complex and their phases may
induce  new contributions to the $CP$ violation, so we write them as
\begin{eqnarray}
\lambda_{ijk}\lambda^*_{lmn} = |\lambda_{ijk}\lambda^*_{lmn}|~e^{i
\phi_{\spur{R}}},~~~~~\lambda^*_{ijk}\lambda_{lmn} =
|\lambda_{ijk}\lambda^*_{lmn}|~e^{-i \phi_{\spur{R}}},
\end{eqnarray}
here $\phi_{\spur{R}}$ is the RPV weak phase, which could be any
value between 0 and $\pi$. To include the effect of $\pi \leq
\phi_{\spur{R}}\leq 2\pi$, $|\lambda_{ijk}\lambda^*_{lmn}|$ is
allowed to take both positive and negative values for simplicity.

 Compared  with the operators in the $\mathcal{H}^{SM}_{eff}$,
there are new operators $(\bar{q}_2q_3)_{V\pm A} (\bar{b}q_1)_{V+A}$
in the $\mathcal{H}^{\spur{R}}$. For $B \to PP$ decays, since
\begin{eqnarray}
&&\langle P|~ \bar{q}_1\gamma_\mu(1-\gamma_{5})q_2|~ 0
\rangle=-\langle P|~ \bar{q}_1\gamma_\mu(1+\gamma_{5})q_2|~ 0
\rangle=-\langle P|~ \bar{q}_1 \gamma_\mu\gamma_{5}q_2|~ 0 \rangle
,\\
&&\langle P|~ \bar{q}\gamma_{\mu}(1-\gamma_{5})b~|~B \rangle
=\langle P|~ \bar{q}\gamma_{\mu}(1+\gamma_{5})b~|~B \rangle=\langle
P|~ \bar{q}\gamma_{\mu}b~|~B \rangle,\hspace{1cm}
\end{eqnarray}
the RPV contribution to the decay amplitude will  modify the SM
amplitude by an overall  relation.

Since we are considering the leading effects of RPV, we need only
evaluate the nonfactorizable vertex corrections and  hard
spectator scattering contributions.
 We ignore the RPV penguin contributions,
 which are expected to be small even compared to the SM penguin
   amplitudes,  due to the smallness of the relevant RPV
couplings compared with the SM gauge couplings. As shown in Ref.
\cite{RPVpipi}, the bounds on the RPV couplings are insensitive to
the inclusion of RPV penguins. We also have neglected the
annihilation contributions in the RPV amplitudes. The R-parity
violating part of the decay amplitudes $\mathcal{A}^{\spur{R}}$  can
be found  in Appendix B.

\subsection{The  branching ratio and the direct $CP$ asymmetries}

With the QCDF approach, we can get the total decay amplitude
\begin{eqnarray}
\mathcal{A}(B\rightarrow P_1 P_2)=\mathcal{A}^{SM}_f(B\rightarrow
P_1 P_2)+\mathcal{A}^{SM}_a(B\rightarrow P_1
P_2)+\mathcal{A}^{\spur{R}}(B\rightarrow P_1 P_2). \label{amp}
\end{eqnarray}
The expressions for the SM amplitude $\mathcal{A}^{SM}_{f,a} $ and
the RPV amplitude $ \mathcal{A}^{\spur{R}}$ are presented in
Appendices A and B, respectively. From the amplitude in Eq.
(\ref{amp}), the branching ratio reads
\begin{eqnarray}
\mathcal{B}r(B\rightarrow P_1 P_2)=\frac{\tau_B |p_c |}{8\pi
m_B^2}\left|\mathcal{A}(B\rightarrow P_1 P_2)\right|^2S,
\end{eqnarray}
where $S=1/2$  for identical $P_1$ and $P_2$,  $S = 1$ otherwise,
$\tau_B$ is the B lifetime, $|p_c|$ is the center of mass momentum
of light mesons in the rest frame of B meson, and given by
\begin{eqnarray}
|p_c|=\frac{1}{2m_B}\sqrt{[m_B^2-(m_{P_1}+m_{P_2})^2][m_B^2-(m_{P_1}-m_{P_2})^2]}.
\end{eqnarray}

The $CP$ averaged branching ratios are defined by
\begin{eqnarray}
\mathcal{B}r(B^{\pm}\to P_1^{\pm}P_2)&\equiv &
\frac{1}{2}\left[\mathcal{B}r(B^{-}\to
P_1^{-}\bar{P}^0_2)+\mathcal{B}r(B^{+}\to P_1^{+}P^0_2)\right],\nonumber\\
\mathcal{B}r(B_d\to P_1^{\pm}P_2^{\mp})&\equiv &
\frac{1}{2}\left[\mathcal{B}r(\bar{B}^{0}_d\to
P_1^{+}P_2^{-})+\mathcal{B}r(B^{0}_d\to P_1^{-}P_2^{+})\right],\\
\mathcal{B}r(B_d\to P_1^0~P_2^0)&\equiv &
\frac{1}{2}\left[\mathcal{B}r(\bar{B}^{0}_d\to
P_1^0\bar{P}_2^0)+\mathcal{B}r(B^{0}_d\to
P_1^0P_2^0)\right].\nonumber
\end{eqnarray}

 The direct  $CP$ asymmetry is defined by
\begin{eqnarray}
\mathcal{A}_{CP}^{dir}=\frac{\mathcal{B}r(\bar{B}\to
\bar{f})-\mathcal{B}r(B\to f)}{\mathcal{B}r(\bar{B}\to
\bar{f})+\mathcal{B}r(B\to f)}.
\end{eqnarray}

\section{Input parameters}

{\bf A. Wilson coefficients} \\
For numerical analyzes,  we use the next-to-leading  Wilson
coefficients calculated in the naive dimensional regularization
(NDR) scheme and at $m_b$ scale
 \cite{coeff}
\begin{eqnarray}
&&C_1=1.082,~~~C_2=-0.185,~~~C_3=0.014,~~~C_4=-0.035,
~~~C_5=0.009,\nonumber\\
&&C_6=-0.041,~~~ C_7/\alpha_e =-0.002,~~~C_8/\alpha_e
=0.054,~~~C_9/\alpha_e
=-1.292, \nonumber \\
&&C_{10}/\alpha_e =0.263,~~~
C^{eff}_{7\gamma}=-0.299,~~~C^{eff}_{8g}=-0.143.\nonumber
\end{eqnarray}
{\bf B. The CKM matrix element} \\
The magnitude of the CKM elements are taken from \cite{PDG2004}
 \begin{equation} \begin{array}{lll}
 |V_{ud}|=0.9738\pm0.0005, & |V_{us}|=0.2200\pm0.0026, & |V_{ub}|=0.00367\pm0.00047,\\
 |V_{cd}|=-0.224\pm0.012, & |V_{cs}|=0.996\pm0.013, & |V_{cb}|=0.0413\pm0.0015,\\
| V_{tb}^*V_{td}|=0.0083\pm0.0016 & |V_{tb}V_{ts}^*|=-0.047\pm0.008, & \\
  \end{array}
 \end{equation}
and the weak  phases $\gamma=60^\circ \pm 14^\circ$,
sin$(2\beta)=0.736\pm 0.049$.\\
{\bf C. Masses and lifetime }\\
 There are two types of quark mass in our analysis. One type is the pole mass which appears
in the loop integration. Here we fix them as
\begin{eqnarray*}
m_u=m_d=m_s=0,~~~~m_c=1.47 ~\mbox{GeV},~~~~m_b=4.8 ~\mbox{GeV}.
\end{eqnarray*}
The other type quark mass appears in the hadronic matrix elements
and the chirally enhanced factor
$r_\chi^P=\frac{2\mu_p}{\overline{m}_b}$ through the equations of
motion, which are renormalization scale dependent. We shall use the
2004 Particle Data Group values \cite{PDG2004} for discussion (the
central values are taken  as our default values)
\begin{eqnarray*}
&&\overline{m}_u(2{\rm GeV})=0.0015\sim 0.004~ {\rm
GeV},~~~~\overline{m}_d(2{\rm GeV})=0.004\sim0.008~ {\rm
GeV},\\
 &&\overline{m}_s(2{\rm GeV})=0.08\sim 0.13 ~{\rm
GeV},~~~~~~~~~\overline{m}_b(\overline{m}_b)=4.1\sim4.4~ {\rm GeV},
\end{eqnarray*}
and then employ the formula in Ref. \cite{coeff}
\begin{eqnarray*}
\overline{m}(\mu)=\overline{m}(\mu_0)\left[\frac{\alpha_s(\mu)}
{\alpha_s(\mu_0)}\right]^{\frac{\gamma^{(0)}_m}{2
\beta_0}}
\left[1+\left(\frac{\gamma^{(1)}_m}{2\beta_0}-\frac{\beta_1\gamma^{(0)}_m}{2\beta_0^2}\right)
\frac{\alpha_s(\mu)-\alpha_s(\mu_0)}{4 \pi}\right]
\end{eqnarray*}
to obtain the current quark masses at $\mu$ scale. The definitions
of $\gamma^{(0)}_m, \gamma^{(1)}_m, \beta_0, \beta_1$ can be found
in
\cite{coeff}.\\
{\bf D. The light cone distribution amplitudes (LCDAs) of the pseudoscalar meson}\\
 For the LCDAs of the
pseudoscalar meson, we use the asymptotic form \cite{M
braun,DAchernyak}
\begin{eqnarray}
\Phi_P(x)=6x(1-x), ~~~~~~~~~~~\Phi^P_p(x)=1.
\end{eqnarray}
We adopt the moments of the $\Phi_1^B (\xi)$ defined in Ref.
\cite{benekeNPB675,BBNS} for our numerical evaluation
\begin{equation}
\int_0^1{\rm d}\xi\frac{\Phi_1^B
(\xi)}{\xi}=\frac{m_{B}}{\lambda_B},
\end{equation}
with $\lambda_B=0.46$ GeV \cite{lamdB}. The quantity $\lambda_B$
parameterizes our ignorance about the $B$ meson distribution
amplitudes and thus brings considerable theoretical  uncertainty.\\
{\bf E. The decay constants and form factors}\\
 For the decay
constants, we take the latest light-cone QCD sum rule (LCSR)
results \cite{BallZwickybtp} in our calculations
\begin{center}
\begin{tabular}{ccc}
$f_{B}=0.162$ GeV,&$f_{K}=0.16$ GeV,&$f_{\pi}=0.131$ GeV.
\end{tabular}
\end{center}
 For the form factors involving  $B\to
K$ and $B\to\pi$ transitions, we adopt the center values of the
results \cite{BallZwickybtp}
 \begin{eqnarray*}
  f^{B\to K}_{0}(0)=0.331, ~~~~~~~~~~~~~~
   f^{B\to \pi}_{0}(0)=0.258.
 \end{eqnarray*}

\section{Numerical results and analysis}
We will present our numerical results in this section. At first, we
will show our estimations in the SM by taking the center value of
the input parameters and compare with the relevant experimental
data. Then, we will consider the RPV effects and  constrain the
relevant RPV couplings by the  averages of Belle and BABAR
measurements of  the $CP$ averaged branching ratios and the direct
CP asymmetries.

When considering  the RPV effects, we will use the input parameters
and the experimental data which are varied randomly within $1\sigma$
level and $2\sigma$ level, respectively.  In the SM, the weak phase
$\gamma$ is well constrained, however,  with the presence of RPV,
this constraint may be relaxed. We would not take $\gamma$ within
the SM range, but vary it randomly in the range of 0 to $\pi$ to
obtain conservative limits on RPV couplings.
 We assume that only one sfermion contributes at one time
with a mass of 100 GeV.  So for other values of the sfermion masses,
the bounds on the couplings in this paper can be easily obtained by
scaling them by factor $\tilde{f}^2\equiv
(\frac{m_{\tilde{f}}}{100GeV})^2$.

The main numerical results in the SM and the relevant data from the
Belle collaborations \cite{babar1,ba2,ba3,ba4,ba5,ba6} and
\textit{BABAR} collaborations\cite{belle1,be2,be3,be4,be5,be6} are
presented in Table I, II and III, which show the results for the
$CP$ averaged branching ratios, the direct $CP$ asymmetries and the
ratios of the $CP$ averaged branching ratios, respectively.

\begin{table}[b]
\centerline{\parbox{16.5cm}{{Table I: The  $CP$ averaged branching
ratios of $B \to \pi\pi, \pi K$  in the SM(in unit of $10^{-6}$).
Experimental data from \textit{BABAR} and Belle and the SM
predictions in the framework of NF and QCDF, where
$\mathcal{B}r^f$ and $\mathcal{B}r^{f+a}$ denote the results
without and with the contributions from weak annihilation,
respectively. }}} \vspace{0.4cm}
\begin{center}
\begin{tabular}
{lccccccc}\hline\hline
 \multicolumn{3}{c@{\hspace{-19.5cm}}}{QCDF} \\
\cline{6-7}\raisebox{2.3ex}[0pt]{Decays}&\raisebox{2.3ex}[0pt]{Belle}
&\raisebox{2.3ex}[0pt]{\textit{BABAR}}&\raisebox{2.3ex}[0pt]{Average}&\raisebox{2.3ex}[0pt]{NF}&
 $\mathcal{B}r^f$&$\mathcal{B}r^{f+a}$ \\ \hline
$B_d \rightarrow \pi^{\pm} \pi^{\mp}$&$4.4\pm0.6\pm0.3$
&$4.7\pm0.6\pm0.2$&$4.56\pm0.46$&7.56&7.88&8.33\\
$B^{\pm} \rightarrow \pi^0 \pi^{\pm}$&$5.0\pm1.2\pm0.5$
&$5.8\pm0.6\pm0.4$&$5.61\pm0.63$&5.19&4.90&4.90\\
$B_d \rightarrow \pi^{0} \pi^{0}$&$2.3^{+0.4+0.2}_{-0.5-0.3}$
&$1.17\pm0.32\pm0.10$&$1.45\pm0.29$&0.17&0.16&0.18\\
$B^{\pm} \rightarrow \pi^{\pm} K$&$22.0\pm1.9\pm1.1$
&$26.0\pm1.3\pm1.0$&$24.57\pm1.31$&12.42&14.77&17.09\\
$B^{\pm}\rightarrow \pi^0 K^{\pm}$&$12.0\pm1.3^{+1.3}_{-0.9}$
&$12.0\pm0.7\pm0.6$&$12.00\pm0.81$&7.23&8.36&9.57\\
$B_d \rightarrow \pi^{\pm} K^{\mp}$&$18.5\pm1.0\pm0.9$
&$17.9\pm0.9\pm0.7$&$18.15\pm0.87$&9.93&11.64&13.54\\
$B_d \rightarrow \pi^0 K$&$11.7\pm2.3^{+1.2}_{-1.3}$
&$11.4\pm0.9\pm0.6$&$11.44\pm1.00$&4.20&5.06&5.97\\
\hline
\end{tabular}
\end{center}
\end{table}
\begin{table}
\centerline{\parbox{15cm}{Table II: The direct $CP$  asymmetries
$\mathcal{A}^{dir}_{CP}$  (in unit of $10^{-2}$)  for  $B \to
\pi\pi, \pi K$. Experimental data from  \textit{BABAR} and Belle.
$\mathcal{A}^{f}_{CP}$ and $\mathcal{A}^{f+a}_{CP}$ are  defined in
the similar way as branching ratios in Table I. }}
\begin{center}
\begin{tabular}
{lcccccc}\hline\hline
 \multicolumn{3}{c@{\hspace{-17cm}}}{QCDF} \\
\cline{5-6}\raisebox{2.3ex}[0pt]{Decays}&\raisebox{2.3ex}[0pt]{Belle}
&\raisebox{2.3ex}[0pt]{\textit{BABAR}}&\raisebox{2.3ex}[0pt]{Average}&
 $\mathcal{A}^f_{CP}$&$\mathcal{A}^{f+a}_{CP}$ \\ \hline
$B_d \rightarrow \pi^{\pm} \pi^{\mp}$&$58\pm15\pm7$
&$9\pm15\pm4$&$31.93\pm11.32$&-5.86&-5.68\\
$B^{\pm} \rightarrow \pi^0 \pi^{\pm}$&$2\pm8\pm1$
&$-1\pm10\pm2$&$0.85\pm6.32$&-0.07&-0.07\\
$B_d \rightarrow \pi^{0} \pi^{0}$&$44^{+53}_{-52}\pm17$
&$12\pm56\pm6$&$28.33\pm39.42$&63.04&60.55\\
$B^{\pm} \rightarrow \pi^{\pm} K$&$5\pm5\pm1$
&$-9\pm5\pm1$&$-2.00\pm3.61$&1.23&1.13\\
$B^{\pm}\rightarrow \pi^0 K^{\pm}$&$4\pm4\pm2$
&$6\pm6\pm1$&$4.70\pm3.60$&8.12&7.29\\
$B_d \rightarrow \pi^{\pm} K^{\mp}$&$-11.3\pm2.2\pm8$
&$-13.3\pm3\pm0.9$&$-12.02\pm1.86$&6.32&5.50\\
$B_d \rightarrow \pi^0 K$&$-12\pm20\pm9$
&$-6\pm18\pm6$&$-8.57\pm14.35$&-2.45&-2.19\\
\hline
\end{tabular}
\end{center}
\end{table}

\begin{table}[h]
\vspace{0.6cm}
 \centerline{\parbox{10.5cm}{Table III: The ratios of
the $CP$ averaged branching ratios.}} \vspace{0.2cm}
\begin{center}
\begin{tabular}
{lcccccc}\hline\hline
 \multicolumn{3}{c@{\hspace{-10cm}}}{QCDF} \\
\cline{4-5}\raisebox{2.3ex}[0pt]{Ratios}&\raisebox{2.3ex}[0pt]{Exp.}
&\raisebox{2.3ex}[0pt]{NF}&
 without ann.& with ann. \\ \hline
$R^{\pi\pi}_{+-}$&$2.30\pm0.35$&$1.28$&$1.16$&1.10\\
$R^{\pi\pi}_{00}$&$0.66\pm0.14$&$0.04$&$0.04$&0.04\\
$R$&$0.79\pm0.06$&$0.80$&$0.79$&0.79\\
$R_c$&$0.98\pm0.08$&$1.16$&$1.13$&1.12\\
$R_n$&$0.79\pm0.08$&$1.18$&$1.15$&1.13\\
\hline
\end{tabular}
\end{center}
\end{table}

 From  Table I, II, III, we can see the puzzle in $B \to
\pi\pi, \pi K$ which we have already mentioned in the
introduction. For example, the new experimental data for $B \to
\pi^0 \pi^0, \pi^0 K^0$ branching ratios are significantly larger
than the SM predictions, moreover, the expected relation
$A^{dir}_{CP}(B^0\to \pi^\pm K^\mp )\approx A^{dir}_{CP}(B^\pm \to
\pi^0 K^\pm )$ obviously contradict to  the experimental data,
even with the opposite sign for them, the value of $R_n$   is
significantly lower than $R_c$, and so on.

Now we turn to the  RPV effects which may give  possible solutions
to the puzzle. We use the $CP$ averaged branching ratios, the direct
$CP$ asymmetries  and the relevant experimental averages of Belle
and {\textit{BABAR} to constrain the  spaces of the RPV parameters.
As known, data on low energy processes can be used to impose rather
strictly constraints   on many of these couplings. The random
variation of the parameters subjecting  to the constraints as
discussed above leads to the scatter plots displayed in  Fig.
\ref{pipi} and \ref{pik}.

\begin{figure}[h]
\begin{center}
\begin{tabular}{cc}
\includegraphics[scale=0.7]{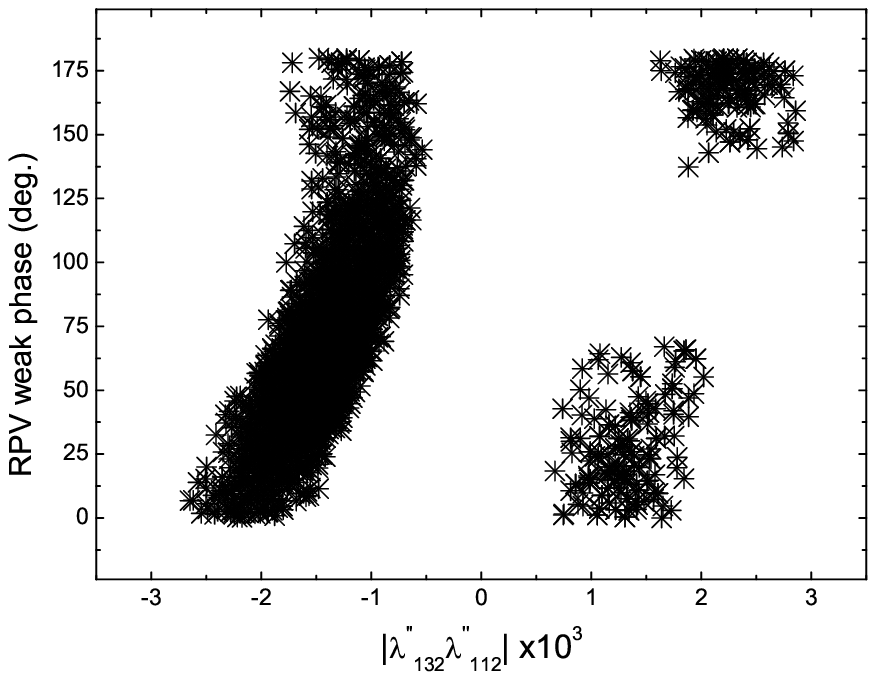}&
\includegraphics[scale=0.7]{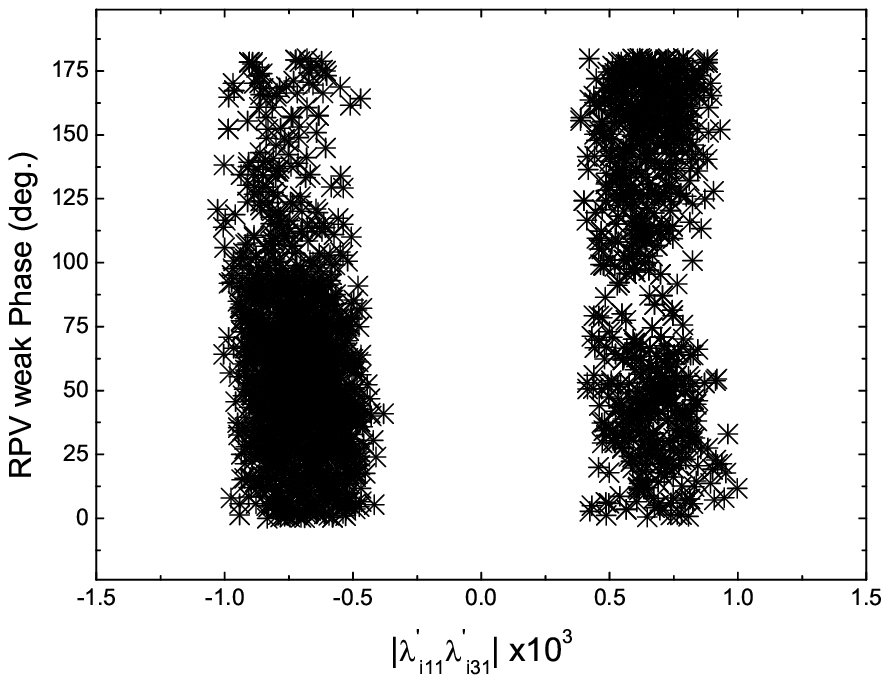}\\
\includegraphics[scale=0.7]{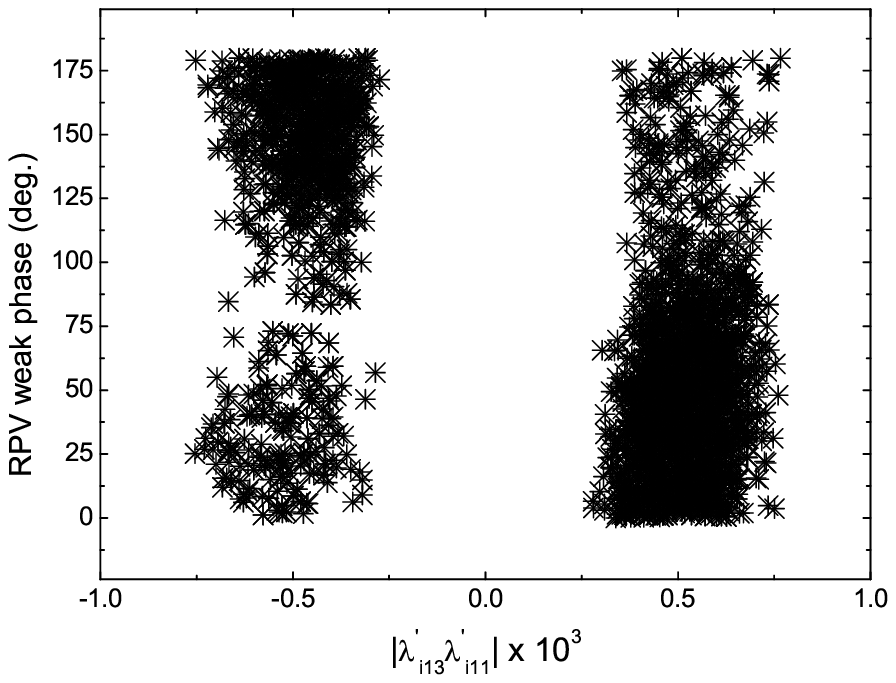}
\end{tabular}
\end{center}
\vspace{-0.6cm}
 \caption{\small The allowed parameter spaces for the
relevant RPV couplings constrained by $B \to \pi\pi$.}
\label{pipi}
\end{figure}

The $B \to \pi\pi$ decays involve the quark level processes
$\bar{b}\to\bar{ d}q \bar{q}~ (q=u,d)$. The  all three RPV couplings
maybe resolve the $B \to \pi\pi$ puzzle, which has been shown in
Fig.\ref{pipi}. These allowed parameter spaces are essentially
controlled by the $CP$ averaged branching ratios and the direct $CP$
asymmetries of different $\pi\pi$ modes. From Fig. \ref{pipi}, we
can see that RPV weak phase is not constrained so much, but the
relevant magnitude of the RPV couplings are constrained within
rather narrow ranges. We can obtain the allowed parameter spaces for
the relevant couplings, which  are summarized in Table IV. For
comparison, we also list the existing bounds on these quadric
coupling products \cite{report,hexg}.

\begin{table}[htb]
\centerline{\parbox{14.5cm}{Table IV: Bounds for  the relevant RPV
coupling constants  by $B \to \pi \pi$ decays for 100 GeV sfermions
and previous bounds  are  listed for comparison. }} \vspace{0.4cm}
\begin{center}
\begin{tabular}{c|c|c|c}\hline\hline
Couplings&Bounds& Process& Previous bounds\\
\hline $|\lambda''_{132}\lambda''_{112}|$&$[5.4\times
10^{-4},~2.9\times 10^{-3}]$ &$B_d\to \pi^{\pm}\pi^{\mp},
\pi^0\pi^0$&$< 5.0\times 10^{-3}$ \cite{report}\\
$|\lambda'_{i13}\lambda'_{i11}|$&$[2.7\times 10^{-4}, ~7.7\times
10^{-4}]$&$B_d\to \pi^{\pm}\pi^{\mp},\pi^0\pi^0$&$<2.5\times
10^{-3}$
\cite{report}\\
$|\lambda'_{i11}\lambda'_{i31}| $&$[3.8\times 10^{-4}, ~1.0\times
10^{-3}]$&$B\to \pi^0\pi^{\pm},\pi^0\pi^0$&$<2.0\times 10^{-3}$
\cite{hexg}\\\hline
\end{tabular}
\end{center}
\end{table}

 We note that since the quark content  of $\pi^0$
is antisymmetric combination $(u\bar{u}-d\bar{d})/\sqrt{2}$, the
decays $ B \to \pi^0 \pi^{\pm},\pi^0\pi^0$ could be induced by
superpartners of both up-type and down-type fermions in RPV SUSY.
For example, $\bar{b}\rightarrow \bar{d}d\bar{d}$ could be induced
by sneutrino, while $\bar{b}\rightarrow \bar{u}u\bar{d}$ could be
induced by slepton with the same $\lambda'_{i13}\lambda'^*_{i11}$
product. We take
$\frac{\lambda'_{i13}\lambda'^*_{i11}}{m^2_{\tilde{e}_i}}$ and
$\frac{\lambda'_{i13}\lambda'^*_{i11}}{m^2_{\tilde{\nu}_i}}$
contribute to $\bar{b}\rightarrow \bar{u}u\bar{d}$ and
$\bar{b}\rightarrow \bar{d}d\bar{d}$ at the same time, and the
effects of $\lambda'_{i13}\lambda'^*_{i11}$ will be summed. So that
$\lambda'_{i13}\lambda'^*_{i11}$ in amplitudes of $B^{\pm} \to
\pi^0\pi^{\pm}$ is  cancelled and in amplitudes of $B_d \to
\pi^0\pi^0$ is partly cancelled  if taking
$m_{\tilde{e}_i}=m_{\tilde{\nu}_i}$.

The $B \to \pi K$ processes are due to $\bar{b}\to \bar{s}q \bar{q}~
(q=u,d)$ transitions at quark level. There are six RPV coupling
constants contributing the  four  $B \to \pi K$ decay modes. We scan
the parameter space for possible solutions, and find that only four
pairs of RPV couplings, $\lambda''_{i31}\lambda''_{i21}$,
$\lambda''_{131}\lambda''_{121}$, $\lambda'_{i13}\lambda'_{i12}$ and
$\lambda'_{i21}\lambda'_{i31}$, can survive  after satisfying all
relevant  experimental data of $B \to \pi K$ decays.  But we do not
get the solutions to  the experimental data at $2\sigma$ level with
the other two pairs of RPV coupling constants,
$\lambda'_{i23}\lambda'_{i11}$ and $\lambda'_{i11}\lambda'_{i32}$.
Figure  \ref{pik} displays the allowed ranges for RPV couplings
which satisfy  all relevant experimental data of the $B \to \pi K $
decays. The constraints for  the four RPV couplings are summarized
in Table V. For comparison, we also list the pervious bounds
\cite{report,hexg}.

\begin{figure}[hb]
\begin{center}
\begin{tabular}{cc}
\includegraphics[scale=0.7]{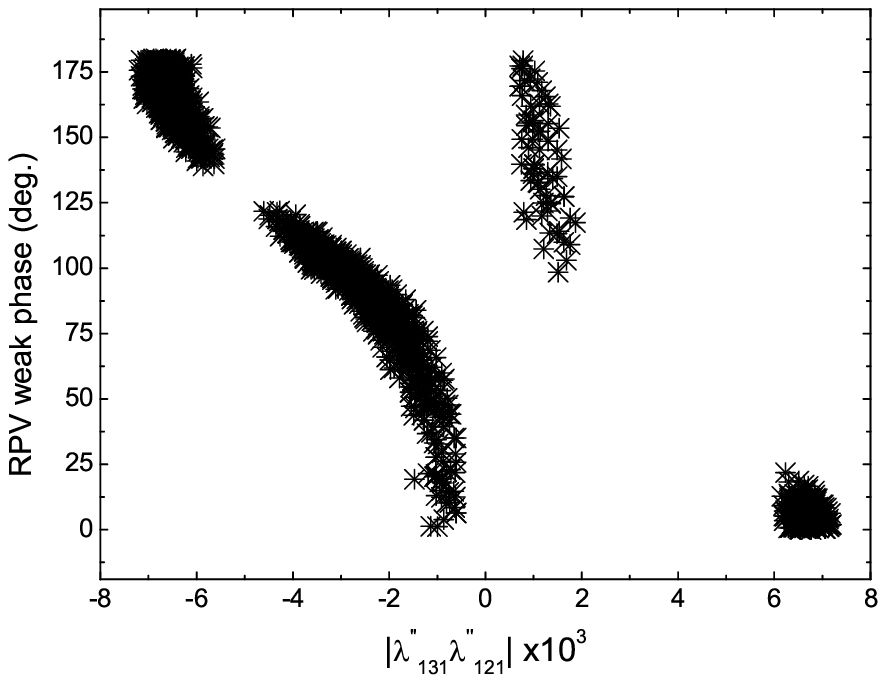}&
\includegraphics[scale=0.7]{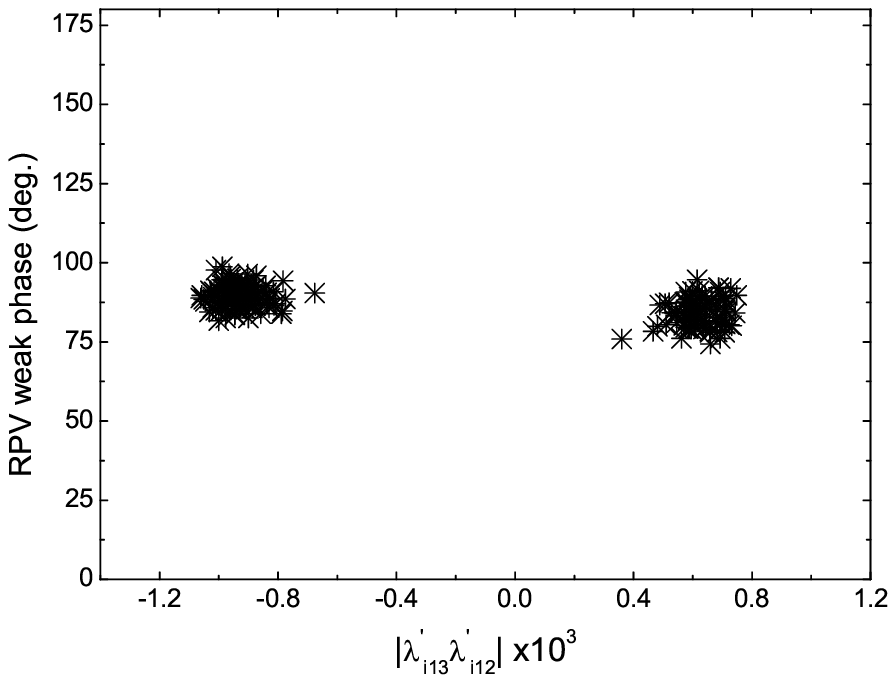}\\
\includegraphics[scale=0.7]{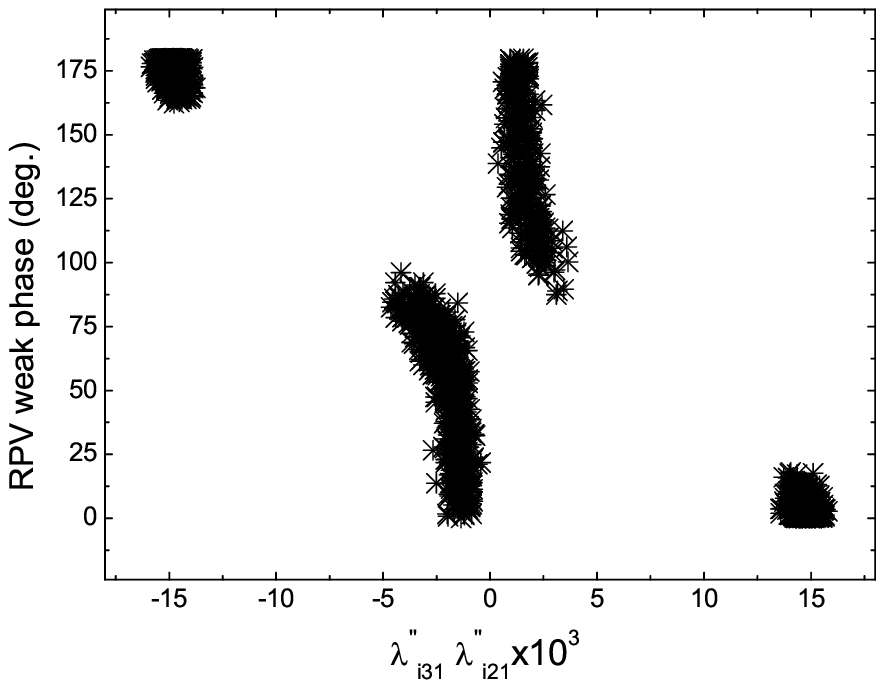}&
\includegraphics[scale=0.7]{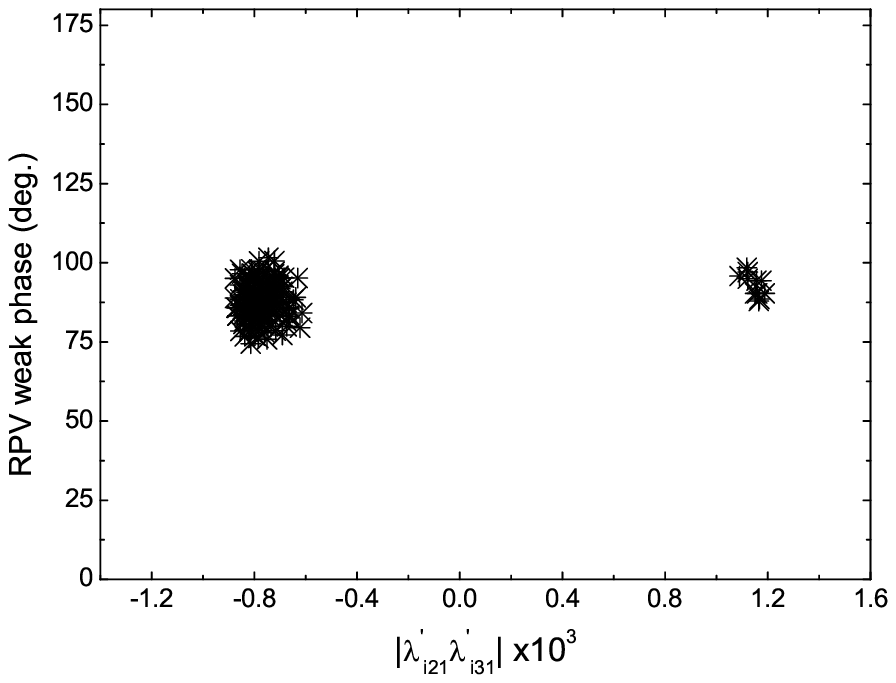}
\end{tabular}
\end{center}
\vspace{-0.6cm}
 \caption{\small The allowed parameter spaces for the
relevant RPV couplings bounded by $B \to \pi K$. We have not listed
the RPV couplings, $\lambda'_{i23}\lambda'_{i11}$ and
$\lambda'_{i11}\lambda'_{i32}$,  for which we have not found the
allowed space.} \label{pik}
\end{figure}

\begin{table}[htbp]
\centerline{\parbox{15.4cm}{Table V: Bounds for  the relevant RPV
coupling couplings for 100 GeV sfermions  by $B \to \pi K$ decays
and previous bounds are also listed. }} \vspace{0.6cm}
\begin{center}
\begin{tabular}{c|c|c|c}\hline\hline
Couplings&Bounds& Process& Previous bounds\\
\hline
$|\lambda''_{i31}\lambda''_{i21}|$&${\displaystyle_{[3.7\times
10^{-4},~4.6\times 10^{-3}]}^{ [13.6\times 10^{-3},~15.8\times
10^{-3}]}}$ &$B^{\pm} \to \pi^{\pm} K, \pi^0 K^{\pm}$&
$<4.\times 10^{-3}\cite{report}$ \\
 $
|\lambda''_{131}\lambda''_{121}|$&${\displaystyle_{[6.1\times
10^{-4},~4.6\times 10^{-3}]}^{ [5.6\times 10^{-3},~7.2\times
10^{-3}]}}$ &$B \to \pi^{0} K^{\pm}, \pi^{\pm}K^{\mp}, \pi^0 K $&
$<4.\times 10^{-3}\cite{report}$ \\
$|\lambda'_{i13}\lambda'_{i12}|$&$ [3.6\times 10^{-4},1.1\times
10^{-3}]$& {\it four modes of } $\pi K$&$<5.7\times 10^{-3}\cite{report}$\\
$|\lambda'_{i21}\lambda'_{i31}|$&$ {\displaystyle_{[6.1\times
10^{-4},~8.7\times 10^{-4}]}^{[1.1\times 10^{-3},~1.2\times
10^{-3}]}}$& $B \to \pi^{\pm}K, \pi^{0} K^{\pm},
 \pi^0 K $&$<1.3\times 10^{-3}\cite{hexg}$\\
\hline
\end{tabular}
\end{center}
\end{table}

For the coupling $\lambda''_{i31}\lambda''_{i21}$, we get two ranges
by the $CP$ averaged branching ratios and the direct $CP$
asymmetries of $B \to \pi K$. The bound for this coupling is
$10^{-1} \sim 10^{-2}$ by branching ratios of $B \to \pi K$ in
\cite{hexg}, and the  limit of $\lambda''_{i3k}\lambda''_{i2k}$ is
0.16 by $B \to K^* \gamma$ decay \cite{report}. So  both spaces of
$\lambda''_{i31}\lambda''_{i21}$  may be allowed.  The
$\lambda''_{131}\lambda''_{121}$ also have two ranges, we may get
only the  space $|\lambda''_{131}\lambda''_{121}|\in [0.61\times
10^{-3},4.60\times 10^{-3}]$ if considering the constraint for
$\lambda''_{i31}\lambda''_{i21}$. The bounds for couplings
$\lambda'_{i13}\lambda'_{i12}$ and $\lambda'_{i21}\lambda'_{i31}$
are obtained  by  four or three  decay modes of $B \to \pi K$, their
ranges are also very narrow. For the RPV couplings
$\lambda'_{i23}\lambda'_{i11}$ and $\lambda'_{i11}\lambda'_{i32}$,
we have not obtained their solutions to the puzzles.

 The $B^{\pm}\to \pi^0 K^{\pm}$ and  $B^0 \to\pi^0
K^0$ decays could be induced by superpartners of both up-type and
down-type fermions, $\bar{b}\rightarrow \bar{d}d\bar{s}$ could be
induced by sneutrino, while $\bar{b}\rightarrow \bar{u}u\bar{s}$
could be induced by slepton with the same
$\lambda'_{i13}\lambda'^*_{i12}$ product. For the same reason as in
$ B \to \pi^0 \pi^{\pm},\pi^0\pi^0$,  the effects of
$\lambda'_{i13}\lambda'^*_{i12}$ have been summed.

The above analysis has shown that   the  puzzles in the $B \to
\pi\pi, \pi K$ decays can be resolved with RPV effects, however,
the solution parameter spaces are always very narrow. The allowed
spaces constrained by the $B \to \pi \pi, \pi K$ decays are
consistent with that by $B \to VV$ decays in our previous study
\cite{BtoVV}.

\section{Conclusions}
The recent observations of $B \to \pi\pi,~\pi K$ decays which are
inconsistent with the SM expectations represent a challenge for
theoretical interpreting. We have employed the QCDF to present a
study of the RPV effects in the $B \to \pi\pi,\pi K$ decays. Our
analysis has shown that a set of RPV couplings play an important
role to resolve the discrepancies between the theoretical
predictions in the SM and the experimental data. However, the
windows of the RPV couplings intervals are found to be always very
narrow. It implies that these couplings, part or all of them,
might be pinned down from the rich experimental phenomena in these
decays. However, it also implies the window could be closed easily
with refined measurements from experiments in the near future.

It should be noted that some of the $\lambda'$ couplings can
generate sizable neutrino masses \cite{barbier,allanch2}. Allanach
{\em et al}. have obtained quite strong upper bound
$\lambda'_{ijj} <10^{-3}$ in the RPV mSUGRA model. Our bounds for
quadric products $|\lambda'_{i13}\lambda'_{i11}|$ and
$|\lambda'_{i11}\lambda'_{i31}|$ are of  order of
$10^{-3}-10^{-4}$. So combining  their constraints from neutrino
masses and ours from $B\to \pi\pi$ decays, the $\lambda'$
resolution window might be closed. However, we note that the
constraints on $\lambda'$ from neutrino masses would depend on the
explicit neutrino masses models with trilinear couplings only,
bilinear couplings only, or both\cite{barbier}.

Furthermore, the $\lambda'_{111}$ coupling has been constrained as
low as   $|\lambda'_{111}|<5.2\times 10^{-4}$ by neutrino-less
double beta decay\cite{2beta}.  There are also  strong bounds
$|\lambda''_{121}|<10^{-4}$ and $|\lambda''_{131}|<10^{-4}$ from
double nucleon decay\cite{msher} and neutron
oscillations\cite{msher,zwirner}, respectively. Combining these
strong bounds, our solutions with $\lambda'_{113}\lambda'_{111}$,
$\lambda'_{111} \lambda'_{131}$ and
$\lambda''_{131}\lambda''_{121} $ RPV products should be excluded.
However, from
  the comprehensive collation of bounds upon  trilinear RPV
  couplings in Ref.\cite{allanach3},
  our other solutions still remain. Explicitly, the $B\to \pi\pi$
  puzzle  could be resolved by the presences of RPV couplings $\lambda''_{132}
  \lambda''_{112}$, $\lambda'_{i11} \lambda'_{i31}$ and
 $\lambda'_{i13} \lambda'_{i11}$ with $i\neq 1$, while the $B\to \pi
 K$ puzzle could be resolved by $\lambda''_{i31}\lambda''_{i21}$
 ($i\neq 1$), $\lambda'_{i13}\lambda'_{i12}$ and
 $\lambda'_{i21}\lambda'_{i31}$.

Generally, we can believe that QCDF calculations for the direct $CP$
asymmetries could be much more accurate than that for the branching
ratios, since many uncertainties could be cancelled in the ratios.
Therefore the constraints from  the direct $CP$ asymmetries would be
more well-founded than those only from branching ratio measurements
\cite{hexg}. Comparing our prediction with the recent experimental
data within $2\sigma$ level about the $CP$ averaged branching ratios
and the direct $CP$ asymmetries,
 we have obtained bounds on the relevant
products of  RPV couplings.
 With more data from BABAR and Belle, one can
significantly shrink the allowed parameter spaces for RPV
couplings. We find that these constraints are consistent with the
previous bounds,
 even most of them are stronger than the existing limits
 \cite{report,BtoVV,hexg}, which may be useful for further
 study of RPV phenomenology.

To summarize, we have shown that the  $B \to \pi \pi $ puzzle and
the $B \to \pi K $ puzzle could be resolved in the  RPV SUSY.
Using the latest experimental data, we get the allowed values of
the relevant RPV couplings, and the most of these new constraints
are stronger than the existing bounds.

\section*{Acknowledgments}
 The work is supported  by National Science
Foundation under contract No.10305003, Henan Provincial Foundation
for Prominent Young Scientists under contract No.0312001700 and
the NCET Program  sponsored by Ministry of Education, China. The
work  of  Y.D  is also supported by Grant No. F01-2004-000-10292-0
of KOSEF-NSFC International Collaborative Research Grant.

{\em Note added:} After this work is finished, we note the
appearance of hep-ph/0509233\cite{pikRPV}, where the $ B \to \pi K$
decays have been carried out by R. Arnowitt {\em et al.} in R-parity
violating and R-parity conserving SUSY model. However, the $B \to
\pi\pi$ modes and the baryon number violating part are not included
in their paper.

\begin{appendix}
 \begin{center}
 {\LARGE{\bf Appendix}}
 \end{center}
 \section{\hspace{-0.6cm}. The amplitudes in the SM}

 The factorized matrix elements are defined by
\begin{eqnarray}
 A_{M_1M_2}&\equiv&\langle M_2|(\bar{q}_1q_2)_{(V-A)}|0\rangle\langle
  M_1|(\bar{q}_3b)_{(V-A)}|B\rangle
  =i (m^2_B-m^2_{M_1})f_0^{B\to
 M_1}(m_{M_2}^2)f_{M_2},\\
 B_{M_1M_2}&\equiv&\langle M_1M_2|(\bar{q}_1q_2)_{(V-A)}|0\rangle\langle
  0|(\bar{q}_3b)_{(V-A)}|B\rangle
  =i f_B f_{M_1}f_{M_2}.
\end{eqnarray}
\begin{eqnarray}
\mathcal{A}^{SM}_f(\bar{B}^0 \rightarrow\pi^+
\pi^-)&=&\frac{G_F}{\sqrt{2}}\left\{\lambda_{ud} a_1+
 \lambda_{pd}\left[ a_4^p+a_{10}^p +r^\pi_\chi(a_6^p+a^p_8)\right]
 \right\}A_{\pi \pi},\\
 \mathcal{A}^{SM}_f(B^- \rightarrow\pi^- \pi^0)&=&\frac{G_F}{2}
 \left\{\lambda_{ud} (a_1+a_2) +
 \frac{3}{2}\lambda_{pd}\left[ -a_7+a_9+a_{10}^p+r^\pi_\chi a^p_8\right]
 \right\}A_{\pi \pi},\\
 \mathcal{A}^{SM}_f(\bar{B}^0 \rightarrow\pi^0 \pi^0)&
 =&-\frac{G_F}{\sqrt{2}}\left\{\lambda_{ud} a_2+
 \lambda_{pd}\left[ -a_4^p+\frac{3}{2}(-a_7+a_9)
 +\frac{1}{2}a^p_{10}-r^\pi_\chi (a^p_6-\frac{1}{2}a_8^p)\right]
 \right\}A_{\pi \pi},\nonumber\\ \\
 \mathcal{A}^{SM}_f(B^- \rightarrow\pi^-
\bar{K}^0)&=&\frac{G_F}{\sqrt{2}}\left\{
 \lambda_{ps}\left[ \left(a_4^p-\frac{1}{2}a_{10}^p\right)
 +r^K_\chi\left(a_6^p-\frac{1}{2}a^p_8\right)\right]
 \right\}A_{\pi K},\\
 \mathcal{A}^{SM}_f(B^- \rightarrow\pi^0 K^-)&=&\frac{G_F}{2}\left\{\lambda_{us} a_1
 +\lambda_{ps}\left[(a_4^p+a^p_{10})+r^K_\chi (a^p_6+a^p_8)
\right]
 \right\}A_{\pi K}\nonumber\\
 &&+\frac{G_F}{2}\left\{\lambda_{us} a_2
 +\frac{3}{2}\lambda_{ps}(-a_7+a_9)
 \right\}A_{K\pi },\\
 \mathcal{A}^{SM}_f(\bar{B}^0 \rightarrow\pi^+ K^-)&
 =&\frac{G_F}{\sqrt{2}}\left\{\lambda_{us} a_1+
 \lambda_{ps}\left[ (a_4^p+a^p_{10})+r^K_\chi (a^p_6+a_8^p)\right]
 \right\}A_{\pi K},\\
 \mathcal{A}^{SM}_f(\bar{B}^0 \rightarrow\pi^0
 \bar{K}^0)&=&-\frac{G_F}{2}\left\{\lambda_{ps}
 \left[ \left(a_4^p-\frac{1}{2}a_{10}^p\right)
 +r^K_\chi\left(a_6^p-\frac{1}{2}a^p_8\right)\right]
 \right\}A_{\pi K}\nonumber\\
 &&+\frac{G_F}{2}\left\{\lambda_{us} a_2
 +\frac{3}{2}\lambda_{ps}(-a_7+a_9)
 \right\}A_{K\pi },\\
 \mathcal{A}^{SM}_a(\bar{B}^0 \rightarrow\pi^+
\pi^-)&=&\frac{G_F}{\sqrt{2}}\left\{\lambda_{ud} b_1+
 \lambda_{pd}\left[ b_3+2 b_4-\frac{1}{2}b^{EW}_3+\frac{1}{2}b^{EW}_4\right]
 \right\}B_{\pi \pi},\\
 \mathcal{A}^{SM}_a(B^- \rightarrow\pi^- \pi^0)&=&0,\\
 \mathcal{A}^{SM}_a(\bar{B}^0 \rightarrow\pi^0 \pi^0)&
 =&\frac{G_F}{\sqrt{2}}\left\{\lambda_{ud} b_1+
 \lambda_{pd}\left[ b_3+2 b_4-\frac{1}{2}b^{EW}_3+\frac{1}{2}b^{EW}_4\right]
 \right\}B_{\pi \pi},\\
 \mathcal{A}^{SM}_a(B^- \rightarrow\pi^-
\bar{K}^0)&=&\frac{G_F}{\sqrt{2}}\left\{\lambda_{us} b_2+
 \lambda_{ps}(b_3+b_3^{EW})
 \right\}B_{\pi K},\\
 \mathcal{A}^{SM}_a(B^- \rightarrow\pi^0 K^-)&=&\frac{G_F}{2}\left\{\lambda_{us} b_2+
 \lambda_{ps}(b_3+b_3^{EW})
 \right\}B_{\pi K},\\
  \mathcal{A}^{SM}_a(\bar{B}^0 \rightarrow\pi^+ K^-)&=&\frac{G_F}{\sqrt{2}}\left\{
 \lambda_{ps}\left(b_3-\frac{1}{2}b_3^{EW}\right)
 \right\}B_{\pi K},\\
 \mathcal{A}^{SM}_a(\bar{B}^0 \rightarrow\pi^0 \bar{K}^0)&=&-\frac{G_F}{2}\left\{
 \lambda_{ps}\left(b_3-\frac{1}{2}b_3^{EW}\right)
 \right\}B_{\pi K},
 \end{eqnarray}
here we note that $\lambda_{ps}=V_{pb}V^*_{ps}$ and
$\lambda_{pd}=V_{pb}V^*_{pd}$.

\section{\hspace{-0.6cm}. The amplitudes for RPV}\vspace{-1cm}
\begin{eqnarray}
\mathcal{A}^{\spur{R}}(\bar{B}^0 \rightarrow\pi^+
\pi^-)&=&\frac{\lambda''^*_{132}\lambda''_{112}}
{8m^2_{\tilde{s}}}\eta^{-4/\beta_0}F_{\pi\pi}A_{\pi\pi}
+\frac{\lambda'^*_{i13}\lambda'_{i11}}{8m^2_{\tilde{e}_{Li}}}\eta^{-8/\beta_0}r^\pi_\chi
A_{\pi\pi},\\
\mathcal{A}^{\spur{R}}(B^- \rightarrow\pi^0
\pi^-)&=&-\left[\frac{\lambda'^*_{i13}\lambda'_{i11}}{8\sqrt{2}m^2_{\tilde{e}_{Li}}}-
\left(\frac{\lambda'^*_{i13}\lambda'_{i11}}{8\sqrt{2}m^2_{\tilde{\nu}_{Li}}}
-\frac{\lambda'^*_{i11}\lambda'_{i31}}{8\sqrt{2}m^2_{\tilde{\nu}_{Li}}}\right)\right]
\eta^{-8/\beta_0}(L_{\pi\pi}-r^\pi_\chi)A_{\pi\pi},\\
\mathcal{A}^{\spur{R}}(\bar{B}^0 \rightarrow\pi^0
\pi^0)&=&\frac{\lambda''^*_{132}\lambda''_{112}}{8m^2_{\tilde{s}}}
\eta^{-4/\beta_0}F_{\pi\pi}A_{\pi\pi}
+\frac{\lambda'^*_{i13}\lambda'_{i11}}{8m^2_{\tilde{e}_{Li}}}
\eta^{-8/\beta_0}L_{\pi\pi}A_{\pi\pi}\nonumber\\
&&-\left(\frac{\lambda'^*_{i13}\lambda'_{i11}}{8m^2_{\tilde{\nu}_{Li}}}
-\frac{\lambda'^*_{i11}\lambda'_{i31}}{8m^2_{\tilde{\nu}_{Li}}}\right)
\eta^{-8/\beta_0}(L_{\pi\pi}-r^\pi_\chi)A_{\pi\pi},\\
\mathcal{A}^{\spur{R}}(B^- \rightarrow\pi^-
\bar{K}^0)&=&\frac{\lambda''^*_{i31}\lambda''_{i21}}
{16m^2_{\tilde{u}_i}}\eta^{-4/\beta_0}F_{\pi
K}A_{\pi
K}+\left(\frac{\lambda'^*_{i13}\lambda'_{i12}}{8m^2_{\tilde{\nu}_{Li}}}
-\frac{\lambda'^*_{i21}\lambda'_{i31}}{8m^2_{\tilde{\nu}_{Li}}}\right)
\eta^{-8/\beta_0}r^K_\chi A_{\pi K}\nonumber\\
&&-\left(\frac{\lambda'^*_{i23}\lambda'_{i11}}{8m^2_{\tilde{\nu}_{Li}}}
-\frac{\lambda'^*_{i11}\lambda'_{i32}}{8m^2_{\tilde{\nu}_{Li}}}\right)\eta^{-8/\beta_0}L_{\pi
K}A_{\pi K},\\
\mathcal{A}^{\spur{R}}(B^- \rightarrow\pi^0
K^-)&=&\frac{\lambda''^*_{131}\lambda''_{121}}
{8\sqrt{2}m^2_{\tilde{d}}}\eta^{-4/\beta_0}(F_{\pi
K}A_{\pi K}-F_{K\pi
}A_{K\pi})+\frac{\lambda''^*_{i31}\lambda''_{i21}}{16\sqrt{2}
m^2_{\tilde{u}_i}}\eta^{-4/\beta_0}F_{K\pi}A_{K\pi}\nonumber\\
&&-\left(\frac{\lambda'^*_{i13}\lambda'_{i12}}{8\sqrt{2}m^2_{\tilde{e}_{Li}}}
-\frac{\lambda'^*_{i13}\lambda'_{i12}}{8\sqrt{2}m^2_{\tilde{\nu}_{Li}}}
\right)\eta^{-8/\beta_0}L_{K\pi}A_{K\pi}\nonumber\\
&&-\left(\frac{\lambda'^*_{i23}\lambda'_{i11}}{8\sqrt{2}m^2_{\tilde{\nu}_{Li}}}
-\frac{\lambda'^*_{i11}\lambda'_{i32}}{8\sqrt{2}m^2_{\tilde{\nu}_{Li}}}
\right)\eta^{-8/\beta_0}r^\pi_\chi
A_{K\pi}\nonumber\\
&&-\frac{\lambda'^*_{i21}\lambda'_{i31}}{8\sqrt{2}m^2_{\tilde{\nu}_{Li}}}
\eta^{-8/\beta_0}L_{K\pi}A_{K\pi}
+\frac{\lambda'^*_{i13}\lambda'_{i12}}{8\sqrt{2}
m^2_{\tilde{e}_{Li}}}\eta^{-8/\beta_0}r^K_\chi
A_{\pi K},\\
\mathcal{A}^{\spur{R}}(\bar{B}^0 \rightarrow\pi^+
K^-)&=&\frac{\lambda''^*_{131}\lambda''_{121}}{8m^2_{\tilde{d}}}\eta^{-4/\beta_0}F_{\pi
K}A_{\pi K}
+\frac{\lambda'^*_{i13}\lambda'_{i12}}{8m^2_{\tilde{e}_{Li}}}\eta^{-8/\beta_0}r^K_\chi
A_{\pi K},\\
\mathcal{A}^{\spur{R}}(\bar{B}^0 \rightarrow\pi^0
\bar{K}^0)&=&-\frac{\lambda''^*_{131}\lambda''_{121}}
{8\sqrt{2}m^2_{\tilde{d}}}\eta^{-4/\beta_0}F_{K\pi}A_{K\pi}
-\frac{\lambda'^*_{i13}\lambda'_{i12}}{8\sqrt{2}
m^2_{\tilde{e}_{Li}}}\eta^{-8/\beta_0}L_{K\pi}
A_{K\pi}\nonumber\\
&&+\left(\frac{\lambda'^*_{i23}\lambda'_{i11}}{8\sqrt{2}m^2_{\tilde{\nu}_{Li}}}
-\frac{\lambda'^*_{i11}\lambda'_{i32}}
{8\sqrt{2}m^2_{\tilde{\nu}_{Li}}}\right)\eta^{-8/\beta_0}
\left(L_{\pi K}A_{\pi K}-r^\pi_\chi
A_{K\pi}\right)\nonumber\\
&&+\left(\frac{\lambda'^*_{i13}\lambda'_{i12}}{8\sqrt{2}m^2_{\tilde{\nu}_{Li}}}
-\frac{\lambda'^*_{i21}\lambda'_{i31}}
{8\sqrt{2}m^2_{\tilde{\nu}_{Li}}}\right)\eta^{-8/\beta_0}
\left(L_{K \pi} A_{K \pi}-r^K_\chi A_{\pi K}\right).
\end{eqnarray}

In the $\mathcal{A}^{\spur{R}}$, $F_{M_1M_2}$ and $A_{M_1M_2}$ are
defined as
\begin{eqnarray}
F_{M_1M_2}&\equiv& 1-\frac{1}{N_C}+\frac{\alpha_s}
{4\pi}\frac{C_F}{N_C}\left(V_{M_2}+H_{M_2M_1}\right),\\
L_{M_1M_2}&\equiv&\frac{1}{N_C}\left[
1-\frac{\alpha_s}{4\pi}\frac{C_F}{N_C}\left(12+V_{M_2}+H_{M_2M_1}\right)\right].
\end{eqnarray}

 \end{appendix}

\end{document}